\documentclass{INTERSPEECH2023}
\usepackage[algo2e,ruled,linesnumbered]{algorithm2e}
\usepackage{multirow}
\usepackage{bbding}
\usepackage{listings}
\usepackage{xcolor} % Required for setting colors

\definecolor{codebg}{rgb}{0.95,0.95,0.95}

\definecolor{codebg}{rgb}{0.96,0.96,0.96} % light gray

\newcommand{\junjie}[1]{{\color{green}#1}}
\interspeechcameraready

\title{WeSep: A Scalable and Flexible Toolkit Towards Generalizable Target Speaker Extraction}
\name{Shuai Wang$^{1,2,5}$, Ke Zhang$^1$, 
Shaoxiong Lin$^3$, Junjie Li$^1$, Xuefei Wang$^1$, Meng Ge$^3$ \\
Jianwei Yu$^4$, Yanmin Qian$^3$, Haizhou Li$^{2,1}$}
\address{\small
  $^1$Shenzhen Research Institute of Big Data, $^2$School of Data Science,  \\
  The Chinese University of Hong Kong, Shenzhen (CUHK-Shenzhen), Guangdong, China \\
  $^3$Auditory Cognition and Computational Acoustics Lab, Shanghai Jiao Tong University, Shanghai, China \\
  $^4$ Tencent AI Lab, Shenzhen,Guangdong, China, 
  $^5$ WeNet Open Source Community}
\email{wangshuai@cuhk.edu.cn}

\begin{document}
\maketitle
 
\begin{abstract}
 Target speaker extraction (TSE) focuses on isolating the speech of a specific target speaker from overlapped multi-talker speech, which is a typical setup in the cocktail party problem. In recent years, TSE draws increasing attention due to its potential for various applications such as user-customized interfaces and hearing aids, or as a crutial front-end processing technologies for subsequential tasks such as speech recognition and speaker recongtion. However, there are currently few open-source toolkits or available pre-trained models for off-the-shelf usage. In this work, we introduce WeSep, a toolkit designed for research and practical applications in TSE. WeSep is featured with flexible target speaker modeling, scalable data management, effective on-the-fly data simulation, structured recipes and deployment support. The toolkit will be publicly avaliable at \url{https://github.com/wenet-e2e/WeSep.}

\end{abstract}

\section{Introduction}

Daily communication environments are often complex, with various audio sources and voices intertwining. Interestingly, humans seem to possess a natural ability: in such complicated backgound, they can effectively focus their attention on the voice of the person they want to listen to. This phonomenon is often termed as ``Selective Attentive Mechnism''~\cite{zmolikova2023neural,bronkhorst2000cocktail,cherry1953some}. Target speaker extraction (TSE) aims to enable a similar process. Unlike blind source separation (BSS), TSE typically relies on additional cue information that directly indicates the identity of the target speaker, thereby circumventing the permutation problem, leading to more flexible and applicable systems. In the current era of large-scale models, it is critical to take advantage of the abundant online media resources. However, before utilizing them for tasks like speech synthesis, it is necessary to process and filter these resources. TSE can play an important role in such pipelines~\cite{yu2023autoprep}.

TSE has gained significant attention in academia and industry. However, the availability of related open-source tools is relatively limited. This scarcity can be attributed to two main factors. Firstly, most TSE research is conducted on synthetic datasets, which may not generalize well to real speech. Secondly, improving the generalization performance for unknown speakers requires advanced speaker modeling techniques. To address these limitations, we aim to provide an accessable open-source toolkit called ``WeSep", focusing on TSE.

The key features of the WeSep toolkit are as follows,
\begin{itemize}
\item To the best of our knowledge, WeSep is the first toolkit focusing on target speaker extraction task, implementing current mainstream models with plans to incorporate more powerful models in the future.

\item WeSep has achieved seamless integration with the open-source speaker modeling toolkit Wespeaker~\cite{wang2023wespeaker}, allowing for flexible integration with powerful pre-trained models and predefined network architectures for joint training.

\item Following the design of WeNet and WeSpeaker, WeSep offers a flexible and efficient data management mechenism called Unified IO (UIO). This mechanism enables WeSep to easily handle large-scale datasets, ensuring scalability and efficiency in data processing.

\item WeSep implements the on-the-fly data simulation pipeline, which allows users to leverage mono-speaker audios prepared for speech recognition or speaker recognition without the need for pre-mixing, thereby enabling model training to scale up and achieve better performance with large datasets.

\item Lastly, models in WeSep can be easily exported by torch Just In Time (JIT) or as the ONNX format, which can be easily adopted in the deployment environment. Pretrained models and sample deployment codes in C++ are also provided.  
\end{itemize}

\section{Related Work}
\subsection{Target Speaker Extraction}
\begin{figure}[ht]
    \centering
    \includegraphics[width=0.5\textwidth]{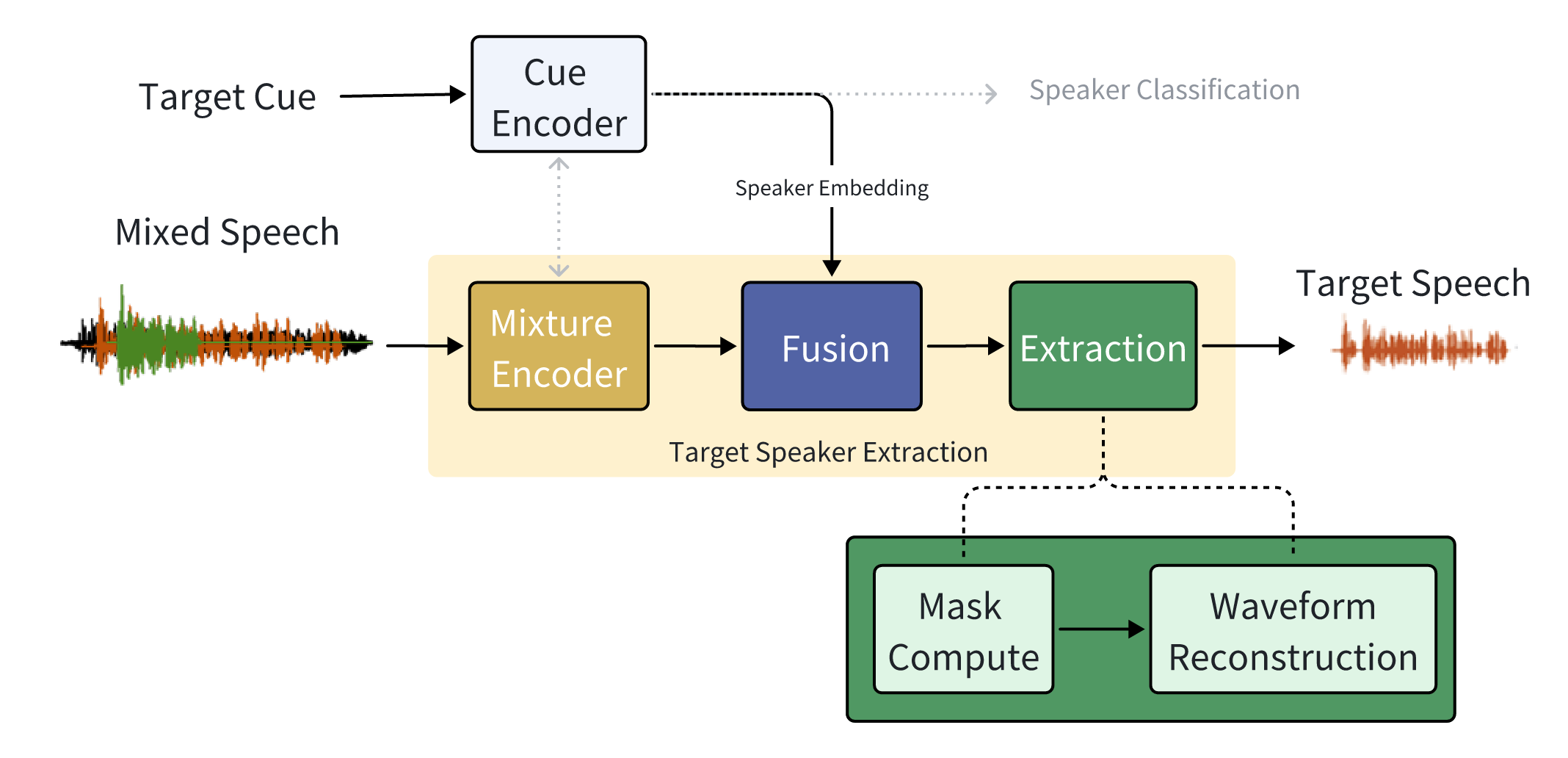}
    \caption{Architecture of a typical TSE system, the cue encoder can be jointly trained or pretrained, an additional speaker classification loss is usually added in the joint-training mode. The parameters of the cue encoder can be shared (or partially shared) with the mixture encoder.}
    \label{fig:tse}
    \vspace{-1em}
\end{figure}

A typical TSE system is depicted in Figure~\ref{fig:tse}.
Assume the mixture signal $m$ containing $K$ speakers is composed of the target speaker $x_s$ and other $K-1$ interfere speakers, as demonstrated by 
\begin{equation}
\label{eq:mix}
    m = x_s + \sum_{k \neq s} ^K x_k + \epsilon
\end{equation}
where $\epsilon$ represents the residual signals capturing noise and reverberation. 

A TSE system aims to reconstruct the $x_s$ from the mixture waveform $m$, given the cue $C_s$. The optimization goal of TSE model $\mathcal{M}_\text{TSE}$ parameterized by $\theta^\text{TSE}$ is to minimize the training loss  $\mathcal{L}(\cdot)$ , which measures how close estimated target speech $\hat{{x}}_s$ is to the target source signal ${x}_s$. 
\begin{align}
\theta^{\text{TSE}}&=\underset{\theta}{\arg \min } \mathcal{L}\left({x}_s, \hat{{x}}_s\right) \\
\hat{{x}}_s&=\mathcal{M}_\text{TSE} \left({m}, {C}_s ; \theta\right)
\end{align}

For audio-based target speaker extraction, the cue $C_s$ typically refers to a pre-enrolled utterance from the target speaker.
% or its processed counterpart, such as a speaker embedding. 
In the case of visual-based TSE\footnote{The visual cue based TSE will be supported in next release}, $C_s$ can be represented by a sequence of image frames capturing the lip movements of the target speaker.

% \vspace{-5mm}
\subsection{Related Open-Source Projects }
% \vspace{-1mm}
Deep learning-based TSE systems have gained popularity in recent years. Although some notable works, such as Spex+\footnote{\url{https://github.com/gemengtju/SpEx_Plus}}~\cite{ge2020spex+} and SpeakerBeam\footnote{\url{https://github.com/BUTSpeechFIT/speakerbeam}}~\cite{vzmolikova2019speakerbeam}, have made their source code publicly available, there is currently no comprehensive toolkit specifically dedicated to this task. Unlike some general-purpose tools like~\cite{watanabe2018espnet} provide simple TSE recipes, WeSep  features a simple code structure that focuses on TSE. In addition to defining speaker model structures within WeSep, users can directly access various state-of-the-art models and pre-trained models from WeSpeaker. 

% \vspace{-3mm}
\section{WeSep}

\subsection{Unified I/O for Local Data Management}
To effectively handle both experimental data and production-scale datasets that encompass tens of thousands of hours of speech, often fragmented into a multitude of small files, we have implemented the Unified Input/Output (UIO) framework~\cite{zhang22g_interspeech} within WeSep. This mechanism has also been integrated into WeNet and WeSpeaker.

\begin{figure*}[!htb]
    \centering
    \includegraphics[width=0.9\textwidth]{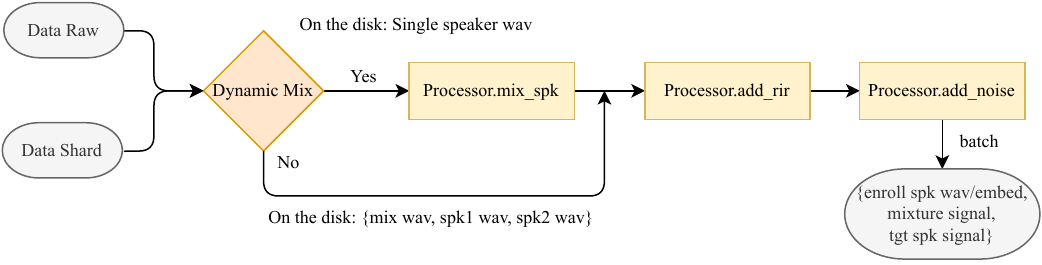}
    \caption{The online data preparation pipeline in WeSep, the case of 2 speakers is demonstrated}
    \label{fig:data_prep}
    \vspace{-1em}
\end{figure*}

\subsection{On-the-Fly Data Simulation}
For datasets like Libri2Mix~\cite{cosentino2020librimix}, researchers typically use pre-processed data and standardized setups to ensure fair comparisons. However, to develop functional systems for real-world applications, it is necessary to train on a substantial amount of data. Preprocessing data and storing it on a hard drive is not an optimal solution. Instead, we propose employing an online data simulation approach as shown in Figure~\ref{fig:data_prep}. This method not only conserves storage resources but also allows for the creation of a more diverse set of training data in a flexible manner, thereby enhancing the robustness of the model. 

\subsubsection{Online Noise and Reverb Generation}

WeSep supports online noise addition and reverberation generation. 
    In line with the approach implemented in WeSpeaker~\cite{wang2023wespeaker}, we draw additive noises from a designated noise database, such as MUSAN~\cite{snyder2015musan} and AudioSet ~\cite{gemmeke2017audio}. However, when it comes to reverberation, WeSep not only offers standard sampling from a Room Impulse Response (RIR) dataset \junjie~\cite{kinoshita2013reverb} but also incorporates the fast random approximation of RIR signals, as introduced in the work by Luo et al.~\cite{luo2023fra}. This enhancement allows for more dynamic and customizable reverberation effects tailored to various acoustic environments.

\subsubsection{Dynamic Speaker Mixing Strategy}
Dynamic Speaker Mixing (DSM) \cite{9495162}involves generating the mixture waveform in real-time during the training process. In contrast to traditional static mixing methods, DSM enhances the model's robustness and generalization ability by introducing greater data diversity and complexity. In WeSep, the DSM algorithm implemented follows Algorithm~\ref{alg:1}.
\begin{algorithm2e}
\footnotesize
\caption{Dynamic Speaker Mixing Strategy}\label{alg:1}
\KwData{ \\
    $n_\text{speaker}$: Number of speakers for the mixed speech\\ 
    $L_\text{Buffer}$: Buffer list containing training utterances\\
    $L_\text{wavs}$: List of wavs to mix \\
    $\text{SNR}_\text{min}$: Min value of SNR (interfere v.s. target speaker)\\
    % $SE_i$: Speaker embedding for speaker $i$ \\
    $\text{SNR}_\text{max}$: Max value of SNR (interfere v.s. target speaker) \\
}
% \KwResult{$y = x^n$}
$L_\text{wavs} \gets [$ $]$\;
\For{$i \gets 0$ \KwTo $n_{speaker}$}{
    \eIf{$i == 0$ }{
        \tcp{Select the utterance for target speaker} 
        
        $s_t \gets random\_sample$($L_\text{Buffer}$);
        
        $L_\text{wavs}$.append($s_t$);
    }{
        \tcp{Select the utterance for interfere speaker and scale with random snr} 
        $snr \gets  random.uniform(\text{SNR}_\text{min},  
  \text{SNR}_\text{max})$;

        $s_i \gets random\_sample(L_\text{Buffer}$);

        \While{same\_speaker$(s_t, s_i)$}{
        $s_i \gets random\_sample(L_\text{Buffer})$;
        }
        $L_\text{wavs}.append(rescale(s_i, snr)$);
    }
    $s_m = add\_and\_rescale(L_\text{wavs})$
}
\textbf{Output:} $s_m$, $L_\text{wavs}$
\end{algorithm2e}

\vspace{-5mm}
\subsection{Backbone Support}
% \subsubsection{Time Domain Models}
\begin{itemize}
    \item \textbf{ConvTasNet}: Proposed in ~\cite{luo2019conv}, ConvTasNet is a pioneering deep learning model for single-channel audio source separation that operates directly in the time domain, utilizing convolutional neural networks to learn and estimate masks for separating target sources from mixtures. Based on Conv-TasNet, WeSep supports its most famous variant talored for the TSE task, Spex+~\cite{ge2020spex+}.
     \item \textbf{BSRNN}: Initially proposed in~\cite{luo2023music} for music source separation, Band-split Recurrent Neural Network (BSRNN) explicitly divides the spectrogram into different frequency bands and performs fine-grained modelling. \cite{yu2023tspeech} adapts BSRNN for the task of personal speech enhancement (PSE) by incorporating an additional speaker embedding, which inspires the implementation of the BSRNN for TSE in WeSep.
     \item \textbf{DPCCN}: The Densely-connected Pyramid Complex Convolutional Network (DPCCN)~\cite{han2022dpccn} is a novel architecture inspired by DenseUNet, incorporating features from Temporal Convolutional Networks (TCNs) and DenseNet to improve separation performance.
     \item \textbf{TF-GridNet}: Proposed in \cite{10094992}, TF-GridNet operates in the T-F domain and stacks several multi-path blocks to leverage local and global spectro-temporal information, representing the State-of-the-art model for speech seperation. In WeSep, speaker embeddings are integrated prior to each multi-path block to specifically tailor it for the TSE task.
 \end{itemize}

\subsection{Target Speaker Modeling}
To guide the extraction of target speaker's speech, a cue $C_s$ is provided, for the audio based TSE, the cue $C_s^a$ is often represented by a fixed-dimensional speaker embedding, extracted from a speaker encoder which is pretrained for the speaker recognition task or jointly trained within the TSE model.
\vspace{-2mm}
\subsubsection{Speaker Encoders}
Besides specific design in well-known architectures, such as the ResNet based speaker encoder in Spex+~\cite{ge2020spex+}, WeSep offers seamless integration with various speaker models that are predefined in WeSpeaker~\cite{wang2023wespeaker}. It provides support for both ``pretrained'' and ``joint training'' modes. The ``pretrained'' mode involves loading the weights released by WeSpeaker\footnote{\url{https://github.com/wenet-e2e/wespeaker/blob/master/docs/pretrained.md}}, while the ``joint training'' mode only requires the model definition to be loaded, with the weights being optimized jointly with the targeted speech enhancement (TSE) task.

\lstset{
    backgroundcolor=\color{codebg},
    frame=single,
    breaklines=true,
    numbers=left,
    basicstyle=\ttfamily\scriptsize
}
\begin{lstlisting}[language=Python]
# psudo-codes for integrating wespeaker models
from wespeaker import get_speaker_model
# TDNN/ECAPA/ResNet/CAM++/...
s = get_speaker_model(spk_model_name)(**spk_args)
m = BSRNN(**sep_args) # Or other backbones
m.speaker_model = s
if use_pretrain_spk_encoder:
    m.spk_model.load_state_dict(pretrain_path)
    m.speaker_model.freeze()
\end{lstlisting}
\vspace{-5mm}

\subsubsection{Fusion methods}
Considering a speaker embedding $\mathbf{e}_s$ derived from the cue $C_s$ and the intermediate outputs $\mathbf{H}={\mathbf{h}_1, \mathbf{h}_2, \cdots, \mathbf{h}_T}$ encoded from the mixed signal $m$, WeSep supports the following fusion methods, both for the pretraining mode and joint training mode.
\begin{itemize}
    \item \textbf{Concat}: Directly replicate $\mathbf{e}_s$ for $T$ times and concatenate it to $\mathbf{H}$, as used in VoiceFilter~\cite{wang2018voicefilter-fpc} and Spex series~\cite{xu2020spex-tjc,ge2020spex+}.
    \item \textbf{Add}: $\mathbf{e}_s$ is first projected to the same dimension with $\mathbf{h}_t$ and do sample-wise addition.
    \item \textbf{Multiply}: $\mathbf{e}_s$ is first projected to the same dimension with $\mathbf{h}_t$ and do sample-wise multiplication. This is adopted mainly in the SpeakerBeam series~\cite{speakerbeam-fjm,delcroix2020improving-tjm}.
    \item \textbf{FiLM}: Feature-wise linear modulation (FiLM)~\cite{perez2018film,cornell2023multi} applies a transformation to  $\mathbf{H}$ by a learned affine transformation, represented by $\mathbf{h}_t' = \gamma(\mathbf{e}_s) \odot \mathbf{h}_t + \beta(\mathbf{e}_s)$, where $\gamma$ and $\beta$ are functions of the speaker embedding $\mathbf{e}_s$, and $\odot$ denotes element-wise multiplication.
\end{itemize}
\vspace{-3mm}
\subsection{Training Strategies}
\subsubsection{Joint Training with Speaker Encoders}
Despite directly leveraging the pretrained speaker encoders for target cue extraction, WeSep also facilitates the joint optimization of the speaker encoder along with other components. An optional speaker classification loss can be easily configured to help contrain the learned speake embedding space.
\vspace{-2mm}
\subsubsection{Online Sampling of the Enrollment}
To improve the model's resilience to varying enrollment conditions, WeSep maintains a correspondence mapping of spk2utt for all training data. This allows for the random selection of an enrollment utterance belonging to the target speaker for each sample, with the optional corruption of noise addition or reverberation effects to simulate more challenging conditions.
\vspace{-2mm}
\subsubsection{Training Objectives}
WeSep follows common TSE research by using negative scale-invariant signal-to-noise ratio (SI-SNR) \cite{vincent2006performance} as the default training objective. For flexibility, we integrated loss functions from Auraloss\footnote{https://github.com/csteinmetz1/auraloss}. Additionally, we implemented GAN-based loss to offer potential enhancement in perceptual quality.

\vspace{-1mm}
\subsection{Deployment}
Models in WeSep can be effortlessly exported to ONNX or PyTorch's Just-In-Time (JIT) format. We provide sample code to facilitate deployment. Additionally, we offer command-line interfaces (CLI) that are accessible through a straightforward ``\textit{pip install}'' process. Users have access to off-the-shelf pretrained models which can be easily used as a standalone tool or for integration into custom pipelines.
\section{Recipes and Results}
WeSep provides recipes for the standard datasets such as Libri2Mix~\cite{cosentino2020librimix}
, following their respective split and pre-mixing strategies. Additionally, WeSep utilizes the VoxCeleb dataset to showcase the construction of a more generalizable TSE system using single-speaker data collected from real-world scenarios. However, due to space limitations, we will focus on a detailed comparison using the Libri2Mix dataset and highlight the generalization capabilities using VoxCeleb. For comprehensive results on other datasets, please refer to the online repository.
\subsection{Libri2Mix}
The performance of different models on the Libri2Mix-Eval dataset are showcased in Table~\ref{tab:libri_results}. In line with the approach detailed in ~\cite{han2022dpccn,ge2020spex+}, we have implemented DPCCN and Spex+ with a default joint training of the speaker encoder. For the remaining models, the ECAPA-TDNN~\cite{desplanques2020ecapa} pretrained on the VoxCeleb2 \cite{chung18b_interspeech} Dev  set by WeSpeaker is utilized.

\begin{table}[!htb]
    \caption{SI-SDR (dB) comparison of different models}
    \centering
    \begin{tabular}{c|c|c|c}
    \toprule
       \multirow{2}{*}{Backbone}   & \multirow{2}{*}{SpkEnc} & \multicolumn{2}{c}{Training Data} \\ \cmidrule{3-4}
      & &  train-100 & train-360 \\\midrule 
       BSRNN  & Pretrain & 13.32 & 16.57 \\
        TF-GridNet  & Pretrain& 12.09 & 15.79 \\\midrule
         DPCCN & Joint Train  & 11.45 & 13.80 \\
        Spex+ & Joint Train & 12.64 & 14.57  \\\bottomrule
    \end{tabular}
    \label{tab:libri_results}
\end{table}

In the sections below, we will provide a detailed analysis of the impact of fusion strategy, speaker model architecture, and the pretrain/joint-train paradigm. Unless otherwise specified, the experiments utilize BSRNN as the default backbone, the pretrained ECAPA-TDNN as the speaker model, multiplication as the default method, and train-100 as the training dataset.

\subsubsection{Impact of the Fusion Strategy}
To incorporate the encoded speaker representation into the TSE system, a fusion mechanism is employed. Four fusion methods are compared in Table~\ref{tab:fusion}, and it is observed that the simple multiplication achieves the best performance, followed by FiLM. Concatenation and addition methods show similar results.

\begin{table}[!htb]
    \caption{Performance comparison of different fusion methods}
    \centering
    \begin{tabular}{c|c|c|c|c}
    \toprule
      Fusion Method   & Concat & Add & Multiply  & FiLM\\\midrule 
        Libri2Mix  & 12.84 & 13.15 & 13.25  &  \textbf{13.32} \\
        AISHELL2Mix  & 4.61 & 5.15 &  4.76 &  \textbf{5.54} \\
\bottomrule
    \end{tabular}
    \label{tab:fusion}
\end{table}

\vspace{-3mm}
\subsubsection{Impact of the Speaker Model}
To assess the compatibility of various pretrained speaker encoders, we present the results of the BSRNN system utilizing different pretrained embeddings in Table~\ref{tab:spk_encoder}. When comparing architectures trained on the same dataset (VoxCeleb2-Dev, 5994 speakers), achieving superior results on the speaker verification task (VoxCeleb1-O) does not necessarily lead to enhanced performance on the TSE task. However, training on a more extensive dataset can lead to improved TSE performance. For instance, the CAM++ model~\cite{wang23ha_interspeech} developed by Alibaba\footnote{\url{https://modelscope.cn/models/iic/speech_campplus_sv_zh-cn_16k-common}}, trained on a dataset of 200,000 Chinese speakers, demonstrates this improvement, despite its poor performance on VoxCeleb1-O, which may be due to the language mismatch.

\begin{table}[!htb]
    \caption{Performance comparison using different pretrained speaker encoders}
    \centering
    \begin{tabular}{c|c|c|c}
    \toprule
      SpkEnc Type   & Train Data  &  SI-SDR (dB) & EER (\%)\\\midrule 
        TDNN~\cite{snyder2018x}  & VoxCeleb2  & 12.41  & 1.721 \\
        ResNet34~\cite{zeinali2019but}  & VoxCeleb2&  13.18 & 0.937 \\
        ECAPA-TDNN~\cite{desplanques2020ecapa} & VoxCeleb2  & 13.32 & 1.072 \\ 
        CAM++~\cite{wang23ha_interspeech} & VoxCeleb2 & 12.29 & 0.845 \\ \midrule
        CAM++ & Ali 200k &  \textcolor{gray}{14.50} & \textcolor{gray}{6.225} \\\bottomrule
    \end{tabular}
    \label{tab:spk_encoder}
    \vspace{-2em}
\end{table}

\subsubsection{Impact of Joint Training}
WeSep facilitates the joint training of the speaker encoder alongside the backbone model. In Table~\ref{tab:joint}, we present some preliminary results of various training paradigms, illustrating that joint training typically yields superior performance\footnote{This doesn't always holds, for instance, joint trained system with CAM++ can not beat the pretrained model with Ali 200k data, further investgation needs to be carried out}. However, we did not observe the anticipated additional performance gain from the inclusion of the speaker classification loss, as suggested in ~\cite{ge2020spex+,delcroix2020improving-tjm}.

\begin{table}[!htb]
    \caption{Joint training v.s. pretrained speaker encoder}
    \centering
    \vspace{-1em}
    \resizebox{0.9\linewidth}{!}{
    \begin{tabular}{c|c|c|c}
    \toprule
      SpkEnc Type   & Joint Training & Multitask & SI-SDR (dB) \\\midrule 
      \multirow{3}{*}{ResNet34} & $\times$ & $\times$  & 13.18 \\
      & \checkmark & $\times$ & 13.96 \\
      & \checkmark & \checkmark & 13.97\\ \midrule
      \multirow{3}{*}{ECAPA-TDNN} & $\times$ & $\times$  & 13.32\\
      & \checkmark & $\times$ & 13.87 \\
      & \checkmark & \checkmark & 13.85\\\bottomrule
    \end{tabular}
    }
    \label{tab:joint}
    \vspace{-2em}
\end{table}

\subsection{VoxCeleb1}
To privide a TSE model with enhanced applicability and to exemplify the training process of such a system utilizing large-scale data, we have offered a Recipe on VoxCeleb1~\cite{nagrani2017voxceleb}.

\begin{table}[!htb]
    \caption{Generalization on out-of-domain dataset}
    \centering
    \begin{tabular}{c|c|c}
    \toprule
    \multirow{2}{*}{Training Dataset} & \multicolumn{2}{c}{SI-SDR (dB)} \\
         & Libri2Mix   & AISHELL2Mix\\\midrule 
        Libri2Mix-train-100   & 13.32 & 5.54 \\
        Libri2Mix-train-360   & 16.57 & 8.17 \\
        % FiLM
        VoxCeleb1 & 16.18  & 10.18\\ \bottomrule
    \end{tabular}
    \label{tab:vox}
\end{table}

As demonstrated in Table~\ref{tab:vox}, the system trained on VoxCeleb1 yields results on Libri2Mix that are comparable to those obtained by the system trained on in-domain data. Moreover, it exhibits significantly better generalization capabilities on AISHELL2Mix.

\section{Conclusion and Future work}

In this paper, we present WeSep, an open-source project focused on Target Speaker Extraction. WeSep is designed with versatile speaker modeling capabilities, enables online data simulation, and offers scalability to large-scale datasets. Looking ahead, WeSep will continually integrate state-of-the-art (SOTA) models, audio-visual recipes, and will expand its capabilities to include blind speech separation tasks within a unified framework.
\section{Acknowledgement}

This work is supported by Internal Project of Shenzhen Research Institute of Big Data under grant No. T00120220002 and No.J00220230014. Thanks to the help from the WeNet Open Source Community, especially Chengdong Liang.

\bibliographystyle{IEEEtran}
\bibliography{mybib}

\begin{thebibliography}{10}
\providecommand{\url}[1]{#1}
\csname url@samestyle\endcsname
\providecommand{\newblock}{\relax}
\providecommand{\bibinfo}[2]{#2}
\providecommand{\BIBentrySTDinterwordspacing}{\spaceskip=0pt\relax}
\providecommand{\BIBentryALTinterwordstretchfactor}{4}
\providecommand{\BIBentryALTinterwordspacing}{\spaceskip=\fontdimen2\font plus
\BIBentryALTinterwordstretchfactor\fontdimen3\font minus \fontdimen4\font\relax}
\providecommand{\BIBforeignlanguage}[2]{{%
\expandafter\ifx\csname l@#1\endcsname\relax
\typeout{** WARNING: IEEEtran.bst: No hyphenation pattern has been}%
\typeout{** loaded for the language `#1'. Using the pattern for}%
\typeout{** the default language instead.}%
\else
\language=\csname l@#1\endcsname
\fi
#2}}
\providecommand{\BIBdecl}{\relax}
\BIBdecl

\bibitem{zmolikova2023neural}
K.~Zmolikova, M.~Delcroix, T.~Ochiai, K.~Kinoshita, J.~{\v{C}}ernock{\`y}, and D.~Yu, ``Neural target speech extraction: An overview,'' \emph{IEEE Signal Processing Magazine}, vol.~40, no.~3, pp. 8--29, 2023.

\bibitem{bronkhorst2000cocktail}
A.~W. Bronkhorst, ``The cocktail party phenomenon: A review of research on speech intelligibility in multiple-talker conditions,'' \emph{Acta acustica united with acustica}, vol.~86, no.~1, pp. 117--128, 2000.

\bibitem{cherry1953some}
E.~C. Cherry, ``Some experiments on the recognition of speech, with one and with two ears,'' \emph{The Journal of the acoustical society of America}, vol.~25, no.~5, pp. 975--979, 1953.

\bibitem{yu2023autoprep}
J.~Yu, H.~Chen, Y.~Bian, X.~Li, Y.~Luo, J.~Tian, M.~Liu, J.~Jiang, and S.~Wang, ``Autoprep: An automatic preprocessing framework for in-the-wild speech data,'' in \emph{ICASSP 2024-2024 IEEE International Conference on Acoustics, Speech and Signal Processing (ICASSP)}, 2024, pp. 1--5.

\bibitem{wang2023wespeaker}
H.~Wang, C.~Liang, S.~Wang, Z.~Chen, B.~Zhang, X.~Xiang, Y.~Deng, and Y.~Qian, ``Wespeaker: A research and production oriented speaker embedding learning toolkit,'' in \emph{ICASSP 2023-2023 IEEE International Conference on Acoustics, Speech and Signal Processing (ICASSP)}, 2023, pp. 1--5.

\bibitem{ge2020spex+}
M.~Ge, C.~Xu, L.~Wang, E.~S. Chng, J.~Dang, and H.~Li, ``{SpEx+: A Complete Time Domain Speaker Extraction Network},'' in \emph{Proc. Interspeech 2020}, 2020, pp. 1406--1410.

\bibitem{vzmolikova2019speakerbeam}
K.~{\v{Z}}mol{\'\i}kov{\'a}, M.~Delcroix, K.~Kinoshita, T.~Ochiai, T.~Nakatani, L.~Burget, and J.~{\v{C}}ernock{\`y}, ``Speakerbeam: Speaker aware neural network for target speaker extraction in speech mixtures,'' \emph{IEEE Journal of Selected Topics in Signal Processing}, vol.~13, no.~4, pp. 800--814, 2019.

\bibitem{watanabe2018espnet}
S.~Watanabe, T.~Hori, S.~Karita, T.~Hayashi, J.~Nishitoba, Y.~Unno, N.~E.~Y. Soplin, J.~Heymann, M.~Wiesner, N.~Chen \emph{et~al.}, ``Espnet: End-to-end speech processing toolkit,'' in \emph{Interspeech}, 2018, pp. 2207--2211.

\bibitem{zhang22g_interspeech}
B.~Zhang, D.~Wu, Z.~Peng, X.~Song, Z.~Yao, H.~Lv, L.~Xie, C.~Yang, F.~Pan, and J.~Niu, ``{WeNet 2.0: More Productive End-to-End Speech Recognition Toolkit},'' in \emph{Proc. Interspeech 2022}, 2022, pp. 1661--1665.

\bibitem{cosentino2020librimix}
J.~Cosentino, M.~Pariente, S.~Cornell, A.~Deleforge, and E.~Vincent, ``Librimix: An open-source dataset for generalizable speech separation,'' \emph{arXiv preprint arXiv:2005.11262}, 2020.

\bibitem{snyder2015musan}
D.~Snyder, G.~Chen, and D.~Povey, ``Musan: A music, speech, and noise corpus,'' \emph{arXiv preprint arXiv:1510.08484}, 2015.

\bibitem{gemmeke2017audio}
J.~F. Gemmeke, D.~P. Ellis, D.~Freedman, A.~Jansen, W.~Lawrence, R.~C. Moore, M.~Plakal, and M.~Ritter, ``Audio set: An ontology and human-labeled dataset for audio events,'' in \emph{2017 IEEE international conference on acoustics, speech and signal processing (ICASSP)}.\hskip 1em plus 0.5em minus 0.4em\relax IEEE, 2017, pp. 776--780.

\bibitem{kinoshita2013reverb}
K.~Kinoshita, M.~Delcroix, T.~Yoshioka, T.~Nakatani, E.~Habets, R.~Haeb-Umbach, V.~Leutnant, A.~Sehr, W.~Kellermann, R.~Maas \emph{et~al.}, ``The reverb challenge: A common evaluation framework for dereverberation and recognition of reverberant speech,'' in \emph{2013 IEEE Workshop on Applications of Signal Processing to Audio and Acoustics}.\hskip 1em plus 0.5em minus 0.4em\relax IEEE, 2013, pp. 1--4.

\bibitem{luo2023fra}
Y.~Luo and J.~Yu, ``{FRA}-{RIR}: Fast random approximation of the image-source method,'' in \emph{Interspeech}, 2023, pp. 3884--3888.

\bibitem{9495162}
N.~Zeghidour and D.~Grangier, ``Wavesplit: End-to-end speech separation by speaker clustering,'' \emph{IEEE/ACM Transactions on Audio, Speech, and Language Processing}, vol.~29, pp. 2840--2849, 2021.

\bibitem{luo2019conv}
Y.~Luo and N.~Mesgarani, ``Conv-tasnet: Surpassing ideal time--frequency magnitude masking for speech separation,'' \emph{IEEE/ACM transactions on audio, speech, and language processing}, vol.~27, no.~8, pp. 1256--1266, 2019.

\bibitem{luo2023music}
Y.~Luo and J.~Yu, ``Music source separation with band-split rnn,'' \emph{IEEE/ACM Transactions on Audio, Speech, and Language Processing}, 2023.

\bibitem{yu2023tspeech}
J.~Yu, H.~Chen, Y.~Luo, R.~Gu, W.~Li, and C.~Weng, ``Tspeech-ai system description to the 5th deep noise suppression (dns) challenge,'' in \emph{ICASSP 2023-2023 IEEE International Conference on Acoustics, Speech and Signal Processing (ICASSP)}, 2023, pp. 1--2.

\bibitem{han2022dpccn}
J.~Han, Y.~Long, L.~Burget, and J.~{\v{C}}ernock{\`y}, ``Dpccn: Densely-connected pyramid complex convolutional network for robust speech separation and extraction,'' in \emph{ICASSP 2022-2022 IEEE International Conference on Acoustics, Speech and Signal Processing (ICASSP)}, 2022, pp. 7292--7296.

\bibitem{10094992}
Z.-Q. Wang, S.~Cornell, S.~Choi, Y.~Lee, B.-Y. Kim, and S.~Watanabe, ``Tf-gridnet: Making time-frequency domain models great again for monaural speaker separation,'' in \emph{ICASSP 2023 - 2023 IEEE International Conference on Acoustics, Speech and Signal Processing (ICASSP)}, 2023, pp. 1--5.

\bibitem{wang2018voicefilter-fpc}
Q.~Wang, H.~Muckenhirn, K.~Wilson, P.~Sridhar, Z.~Wu, J.~Hershey, R.~A. Saurous, R.~J. Weiss, Y.~Jia, and I.~L. Moreno, ``Voicefilter: Targeted voice separation by speaker-conditioned spectrogram masking,'' in \emph{Interspeech}, 2018.

\bibitem{xu2020spex-tjc}
C.~Xu, W.~Rao, E.~S. Chng, and H.~Li, ``Spex: Multi-scale time domain speaker extraction network,'' \emph{IEEE/ACM transactions on audio, speech, and language processing}, vol.~28, pp. 1370--1384, 2020.

\bibitem{speakerbeam-fjm}
M.~Delcroix, K.~Zmolikova, K.~Kinoshita, A.~Ogawa, and T.~Nakatani, ``Single channel target speaker extraction and recognition with speaker beam,'' in \emph{2018 IEEE international conference on acoustics, speech and signal processing (ICASSP)}, 2018, pp. 5554--5558.

\bibitem{delcroix2020improving-tjm}
M.~Delcroix, T.~Ochiai, K.~Zmolikova, K.~Kinoshita, N.~Tawara, T.~Nakatani, and S.~Araki, ``Improving speaker discrimination of target speech extraction with time-domain speakerbeam,'' in \emph{ICASSP 2020-2020 IEEE International Conference on Acoustics, Speech and Signal Processing (ICASSP)}, 2020, pp. 691--695.

\bibitem{perez2018film}
E.~Perez, F.~Strub, H.~De~Vries, V.~Dumoulin, and A.~Courville, ``Film: Visual reasoning with a general conditioning layer,'' in \emph{Proceedings of the AAAI conference on artificial intelligence}, vol.~32, no.~1, 2018.

\bibitem{cornell2023multi}
S.~Cornell, Z.-Q. Wang, Y.~Masuyama, S.~Watanabe, M.~Pariente, and N.~Ono, ``Multi-channel target speaker extraction with refinement: The wavlab submission to the second clarity enhancement challenge,'' \emph{arXiv preprint arXiv:2302.07928}, 2023.

\bibitem{vincent2006performance}
E.~Vincent, R.~Gribonval, and C.~F{\'e}votte, ``Performance measurement in blind audio source separation,'' \emph{IEEE transactions on audio, speech, and language processing}, vol.~14, no.~4, pp. 1462--1469, 2006.

\bibitem{desplanques2020ecapa}
B.~Desplanques, J.~Thienpondt, and K.~Demuynck, ``Ecapa-tdnn: Emphasized channel attention, propagation and aggregation in tdnn based speaker verification,'' \emph{arXiv preprint arXiv:2005.07143}, 2020.

\bibitem{chung18b_interspeech}
J.~S. Chung, A.~Nagrani, and A.~Zisserman, ``{VoxCeleb2: Deep Speaker Recognition},'' in \emph{Proc. Interspeech 2018}, 2018, pp. 1086--1090.

\bibitem{wang23ha_interspeech}
H.~Wang, S.~Zheng, Y.~Chen, L.~Cheng, and Q.~Chen, ``{CAM++: A Fast and Efficient Network for Speaker Verification Using Context-Aware Masking},'' in \emph{Proc. INTERSPEECH 2023}, 2023, pp. 5301--5305.

\bibitem{snyder2018x}
D.~Snyder, D.~Garcia-Romero, G.~Sell, D.~Povey, and S.~Khudanpur, ``X-vectors: Robust dnn embeddings for speaker recognition,'' in \emph{2018 IEEE international conference on acoustics, speech and signal processing (ICASSP)}.\hskip 1em plus 0.5em minus 0.4em\relax IEEE, 2018, pp. 5329--5333.

\bibitem{zeinali2019but}
H.~Zeinali, S.~Wang, A.~Silnova, P.~Mat{\v{e}}jka, and O.~Plchot, ``But system description to voxceleb speaker recognition challenge 2019,'' \emph{arXiv preprint arXiv:1910.12592}, 2019.

\bibitem{nagrani2017voxceleb}
A.~Nagrani, J.~S. Chung, and A.~Zisserman, ``Voxceleb: a large-scale speaker identification dataset,'' \emph{Telephony}, vol.~3, pp. 33--039, 2017.

\end{thebibliography}

\end{document}